\documentclass[prd,aps,groupedaddress,showpacs]{revtex4}
\usepackage{epsf,epsfig,graphicx}

\newcounter{muni}

\begin{document}
\hbadness=10000 \pagenumbering{arabic}

\title{Implication of the $B\to\rho\rho$ data on the $B\to \pi \pi$ puzzle}

\author{Hsiang-nan Li$^{1}$}
\email{hnli@phys.sinica.edu.tw}
\author{Satoshi Mishima$^2$}
\email{mishima@ias.edu}

\affiliation{$^{1}$Institute of Physics, Academia Sinica, Taipei,
Taiwan 115, Republic of China,} \affiliation{$^{1}$Department of
Physics, National Cheng-Kung University, Tainan, Taiwan 701,
Republic of China}

\affiliation{$^{2}$School of Natural Sciences, Institute for
Advanced Study, Princeton, NJ 08540, U.S.A.}

\begin{abstract}

We point out that the $B\to\rho\rho$ data have seriously
constrained the possibility of resolving the $B\to\pi\pi$ puzzle
from the large observed $B^0\to\pi^0\pi^0$ branching ratio in the
available theoretical approaches. The next-to-leading-order (NLO)
contributions from the vertex corrections, the quark loops, and
the magnetic penguin evaluated in the perturbative QCD (PQCD)
approach have saturated the experimental upper bound of the
$B^0\to\rho^0\rho^0$ branching ratio, and do not help. The NLO
PQCD predictions for the $B^0\to\rho^\mp\rho^\pm$ and
$B^\pm\to\rho^\pm\rho^0$ branching ratios are consistent with the
data. The inclusion of the NLO jet function from the
soft-collinear effective theory into the QCD-improved
factorization approach, though enhancing the $B^0\to\pi^0\pi^0$
branching ratio sufficiently, overshoots the bound of the
$B^0\to\rho^0\rho^0$ branching ratio, and deteriorates the
predictions for the $B^\pm\to \pi^0 K^\pm$ and $B^0\to \pi^\mp
K^\pm$ direct CP asymmetries.

\end{abstract}


\maketitle

\section{INTRODUCTION}

The observed direct CP asymmetries and branching ratios of the
$B\to \pi K$, $\pi\pi$ decays \cite{HFAG},
\begin{eqnarray}
A_{CP}(B^0\to \pi^\mp K^\pm)&=&(-10.8\pm 1.7)\%\;,\nonumber\\
A_{CP}(B^\pm\to \pi^0 K^\pm)&=&(4\pm 4)\%\;,\nonumber\\
B(B^0\to\pi^\mp\pi^\pm)&=&(4.9\pm 0.4)\times 10^{-6}\;,\nonumber\\
B(B^0\to\pi^0\pi^0)&=&(1.45\pm 0.29)\times 10^{-6}\;,\label{data}
\end{eqnarray}
were regarded as puzzles, since they obviously contradict to the
expected relations $A_{CP}(B^0\to \pi^\mp K^\pm)\approx
A_{CP}(B^\pm\to \pi^0 K^\pm)$ and $B(B^0\to\pi^\mp\pi^\pm)\gg
B(B^0\to\pi^0\pi^0)$. These puzzles have been analyzed in the
perturbative QCD (PQCD) approach \cite{KLS,LUY} up to
next-to-leading-order (NLO) accuracy recently \cite{LMS05}, where
the contributions from the vertex corrections, the quark loops,
and the magnetic penguin were taken into account. It was found
that the vertex corrections modify the color-suppressed tree
contribution, such that the relative strong phase between the tree
and penguin amplitudes involved in the $B\to\pi K$ decays
decreases. The predicted magnitude of the $B^\pm \to\pi^0 K^\pm$
direct CP asymmetry then becomes smaller, and matches the data in
Eq.~(\ref{data}). Though the $B\to\pi K$ puzzle has been resolved,
the $B\to\pi\pi$ puzzle remains, because the NLO color-suppressed
tree amplitude does not increase the predicted $B^0\to\pi^0\pi^0$
branching ratio sufficiently.

A resolution to a puzzle usually demands an introduction of new
mechanism. It is thus essential to investigate whether the
proposed new mechanism deteriorates the consistency of theoretical
results with other data. To make sure the above NLO effects are
reasonable, we apply the same PQCD formalism to more two-body
nonleptonic $B$ meson decays, concentrating on the $B\to\rho\rho$
branching ratios, which are also sensitive to the color-suppressed
tree contribution. It will be shown that the NLO PQCD predictions
are in agreement with the data of the $B^0\to\rho^\mp\rho^\pm$ and
$B^\pm\to\rho^\pm\rho^0$ branching ratios, and saturate the
experimental upper bound of the $B^0\to\rho^0\rho^0$ branching
ratio, $B(B^0\to\rho^0\rho^0)<1.1\times 10^{-6}$ \cite{HFAG}.
Therefore, our resolution to the $B\to\pi K$ puzzle makes sense,
and the $B\to\pi\pi$ puzzle is confirmed. The dramatic difference
between the $B\to\pi\pi$ and $\rho\rho$ data has been also noticed
in \cite{BLS0602}, which stimulates the proposal of a new isospin
amplitude with $I = 5/2$. The possible new physics signals from
the $B\to\pi\pi$ decays have been discussed in
\cite{BBLS,YWL2,CGHW}.

It has been claimed that the $B\to\pi\pi$ puzzle is resolved in
the QCD-improved factorization (QCDF) approach \cite{BBNS} with an
input from soft-collinear effective theory (SCET) \cite{BY05}: the
inclusion of the NLO jet function, one of the hard coefficients of
SCET$_{\rm II}$, into the QCDF formula for the color-suppressed
tree amplitude leads to enough enhancement of the
$B^0\to\pi^0\pi^0$ branching ratio. Following the argument made
above, we apply the same formalism \cite{BY05} to the $B\to\pi K$,
$\rho\rho$ decays as a check. It turns out that the effect of the
NLO jet function deteriorates the QCDF results for the direct CP
asymmetries in the $B^\pm\to \pi^0 K^\pm$ and $B^0\to \pi^\mp
K^\pm$ decays: the magnitude of the former increases, while that
of the latter decreases, contrary to the tendency indicated by the
data. This NLO effect also overshoots the upper bound of the
$B^0\to\rho^0\rho^0$ branching ratio very much. This observation
is expected: the $B^0\to\rho^0\rho^0$ and $B^0\to\pi^0\pi^0$
decays have the similar factorization formulas, so the branching
ratio $B(B^0\to\rho^0\rho^0)$ ought to be larger than
$B(B^0\to\pi^0\pi^0)$ due to the meson decay constants $f_\rho >
f_\pi$. Therefore, the $B\to\rho\rho$ data have seriously
constrained the possibility of resolving the $B\to\pi\pi$ puzzle
in the available theoretical approaches.

There exists an alternative phenomenological application of SCET
\cite{BPRS,BPS05}, where the jet function, characterized by the
scale of $O(\sqrt{m_b\Lambda})$, $m_b$ being the $b$ quark mass
and $\Lambda$ a hadronic scale, is regarded as being incalculable.
Its contribution, together with other nonperturbative parameters,
such as the charming penguin, were then determined by the
$B\to\pi\pi$ data. That is, the color-suppressed tree amplitude
can not be explained, but the data are used to fit for the
phenomenological parameters in the theory. Predictions for the
$B\to\pi K$, $KK$ decays were then made based on the obtained
parameters and partial SU(3) flavor symmetry \cite{BPS05}.
Final-state interaction (FSI) is certainly a plausible resolution
to the $B\to\pi\pi$ puzzle, but the estimate of its effect is
quite model-dependent. Even opposite conclusions were drawn
sometimes. When including FSI either into naive factorization
\cite{CHY} or into QCDF \cite{CCS}, the $B^0\to\pi^0\pi^0$
branching ratio was treated as an input in order to fix the
involved free parameters. Hence, no resolution was really
proposed. It has been found that FSI, evaluated in the Regge
model, is insufficient to account for the observed
$B^0\to\pi^0\pi^0$ branching ratio \cite{DLLN}. We conclude that
there is no satisfactory resolution in the literature: the
available proposals are either data fitting, or can not survive
the constraints from the $B\to\pi K$, $\rho\rho$ data under the
current theoretical development.

In Sec.~II we compute the branching ratios, the direct CP
asymmetries, and the polarization fractions of the $B\to\rho\rho$
decays using the NLO PQCD formalism. The branching ratios and the
direct CP asymmetries of the $B\to\pi K$, $\pi\pi$, $\rho\rho$
decays are calculated in Sec.~III by including the NLO jet
function from SCET$_{\rm II}$ into the QCDF formulas. Section IV
is the discussion, where we comment on and compare the various
analyses of the FSI effects in the $B\to\pi K$, $\pi\pi$ decays.

\section{$B\to\rho\rho$ IN NLO PQCD}

The NLO contributions from the vertex corrections, the quark
loops, and the magnetic penguin to the $B\to\pi K$ and $\pi\pi$
decays have been calculated in the naive dimensional
regularization (NDR) scheme in the PQCD approach \cite{LMS05}, and
the results for the branching ratios and the direct CP asymmetries
are quoted in Tables~\ref{br1} and \ref{cp1}, respectively. We
have taken this chance to correct a minor numerical mistake in the
vertex corrections for the $B\to\pi K$ decays, whose branching
ratios become smaller by $2\sim 4\%$. Note that a minus sign is
missing for the $q=t$ term in the expression for the quark-loop
contributions in Eq.~(27) of \cite{LMS05}. Nevertheless, this typo
has nothing to do with the numerical outcomes. Our observations
are summarized below. The corrections from the quark loops and
from the magnetic penguin come with opposite signs, and sum to
about $-10\%$ of the leading-order (LO) penguin amplitudes. They
mainly reduce the penguin-dominated $B\to\pi K$ branching ratios,
but have a minor influence on the tree-dominated $B\to\pi\pi$
branching ratios, and on the direct CP asymmetries. On the
contrary, the vertex corrections do not change the branching
ratios, except the $B^0\to\pi^0\pi^0$ one. They modify only the
direct CP asymmetries of the $B^\pm\to\pi^0 K^\pm$,
$B^0\to\pi^0K^0$, and $B^0\to\pi^0\pi^0$ modes by increasing the
color-suppressed tree amplitude $C'$ few times. The larger $C'$,
leading to the nearly vanishing direct CP asymmetry
$A_{CP}(B^\pm\to \pi^0K^\pm)$, resolves the $B\to\pi K$ puzzle
within the standard model.

\begin{table}[hbt]
\begin{center}
\begin{tabular}{cccccccc}
\hline\hline Mode & Data \cite{HFAG}& LO & LO$_{\rm NLOWC}$ & +VC
& +QL &  +MP  & +NLO
\\
\hline $B^\pm \to \pi^\pm K^0$ & $ 24.1 \pm 1.3 $ &
 $17.0$&$32.3$&$30.1$&$34.2$&$24.1$&
 $23.6^{+14.5\,(+13.8)}_{-\ 8.4\,(-\ 8.2)}$
\\
$B^\pm \to \pi^0 K^\pm$ & $ 12.1 \pm 0.8 $ &
 $10.2$&$18.4$&$17.1$&$19.4$&$14.0$&
 $13.6^{+10.3\,(+\ 7.3)}_{-\ 5.7\,(-\ 4.3)}$
\\
$B^0 \to \pi^\mp K^\pm$ & $ 18.9 \pm 0.7 $ &
 $14.2$&$27.7$&$26.1$&$29.4$&$20.5$&
 $20.4^{+16.1\,(+11.5)}_{-\ 8.4\,(-\ 6.7)}$
\\
$B^0 \to \pi^0 K^0 $ & $ 11.5 \pm 1.0 $ &
 $\phantom{0}5.7$&$12.1$&$11.4$&$12.8$&$\phantom{0}8.7$&
 $\phantom{0}8.7^{+\ 6.0\,(+\ 5.5)}_{-\ 3.4\,(-\ 3.1)}$
\\
\hline $B^0 \to \pi^\mp \pi^\pm$ & $ \phantom{0}4.9 \pm 0.4 $ &
 $\phantom{0}7.0$&$\phantom{0}6.8$&$\phantom{0}6.6$&
 $\phantom{0}6.9$&$\phantom{0}6.7$&
 $\phantom{0}6.5^{+\ 6.7\,(+\ 2.7)}_{-\ 3.8\,(-\ 1.8)}$
\\
$B^\pm \to \pi^\pm \pi^0$ & $ \phantom{0}5.5 \pm 0.6 $ &
 $\phantom{0}3.5$&$\phantom{0}4.1$&$\phantom{0}4.0$&
 $\phantom{0}4.1$&$\phantom{0}4.1$&
 $\phantom{0}4.0^{+\ 3.4\,(+\ 1.7)}_{-\ 1.9\,(-\ 1.2)}$
\\
$B^0 \to \pi^0 \pi^0$ & $ \phantom{0}1.45 \pm 0.29 $ &
 $\phantom{0}0.12$&$\phantom{0}0.27$&$\phantom{0}0.37$&
 $\phantom{0}0.29$&$\phantom{0}0.21$&
 $\phantom{0}0.29^{+0.50\,(+0.13)}_{-0.20\,(-0.08)}$
\\
\hline\hline
\end{tabular}
\end{center}
\caption{Branching ratios from PQCD in the NDR scheme in units of
$10^{-6}$. The label LO$_{\rm NLOWC}$ means the LO results with
the NLO Wilson coefficients, and +VC, +QL, +MP, and +NLO mean the
inclusions of the vertex corrections, of the quark loops, of the
magnetic penguin, and of all the above NLO corrections,
respectively. The errors in the parentheses represent only the
hadronic uncertainty \cite{LMS05}.}\label{br1}
\end{table}

\begin{table}[hbt]
\begin{center}
\begin{tabular}{cccccccc}
\hline\hline Mode & Data \cite{HFAG}& LO & LO$_{\rm NLOWC}$& +VC &
+QL &  +MP  &  +NLO
\\
\hline $B^\pm \to \pi^\pm K^0$ & $ -2 \pm  4$ &
 $\ \, -1$&$-1$&$-1$& $\phantom{-}0$&$\ \, -1$&
 $\ \ \ \ \, \phantom{-}0\pm 0\,(\pm 0)$
\\
$B^\pm \to \pi^0 K^\pm$ & $ \phantom{-}4 \pm 4 $ &
 $\ \, -8$&$-6$&$-2$&$-5$&$\ \, -8$&
 $\ \, -1^{+3\,(+3)}_{-6\,(-5)}$
\\
$B^0 \to \pi^\mp K^\pm$ & $ -10.8 \pm 1.7 $ &
 $-12$&$-8$&$-9$&$-6$&$-10$&
 $-10^{+7\,(+5)}_{-8\,(-6)}$
\\
$B^0 \to \pi^0 K^0 $ & $ \phantom{-}2 \pm 13 $ &
 $\ \, -2$& $\phantom{-}0$&$-7$& $\phantom{-}0$&
 $\ \, \phantom{-}0$&
 $\ \, -7^{+3\,(+1)}_{-4\,(-2)}$
\\
\hline $B^0 \to \pi^\mp \pi^\pm$ & $ \phantom{-}37 \pm 10$ &
 $\phantom{-}14$& $\phantom{-}19$& $\phantom{-}21$&
 $\phantom{-}16$& $\phantom{-}20$&
 $\phantom{-}18^{+20\,(+\ 7)}_{-12\,(-\ 6)}$
\\
$B^\pm \to \pi^\pm \pi^0$ & $ \phantom{-}1 \pm 6 $&
 $\ \, \phantom{-}0$& $\ \, \phantom{-}0$& $\ \, \phantom{-}0$&
 $\ \, \phantom{-}0$& $\ \, \phantom{-}0$&
 $\ \ \ \ 0\pm 0\,(\pm 0)$
\\
$B^0 \to \pi^0 \pi^0$ & $ \phantom{-}28^{+40}_{-39}$ &
 $\ \, -4$&$-34$& $\phantom{-}65$&$-41$&$-43$&
 $\phantom{-}63^{+35\,(+\ 9)}_{-34\,(-15)}$
\\
\hline\hline
\end{tabular}
\end{center}
\caption{Direct CP asymmetries from PQCD in the NDR scheme in
percentage.}\label{cp1}
\end{table}

The above observations can be easily understood as follows. The
$B^0\to \pi^\mp K^\pm$ decays involve the color-allowed tree $T'$
and the QCD penguin $P'$ in the topological amplitude
parametrization. The data of $A_{CP}(B^0\to \pi^\mp K^\pm)\approx
-11.5\%$ imply a sizable relative strong phase between $T'$ and
$P'$. The $B^\pm\to \pi^0 K^\pm$ decays involve $C'$ and the
electroweak penguin amplitude $P'_{ew}$, in addition to $T'$ and
$P'$. If $C'$ is large enough, and more or less orthogonal to
$T'$, it may orient the sum $T'+C'$ roughly along with
$P'+P'_{ew}$. The smaller relative strong phase between $T'+C'$
and $P'+P'_{ew}$ then gives $A_{CP}(B^\pm\to \pi^0 K^\pm)\approx
0$. We found in PQCD that the vertex corrections indeed modify
$C'$ in this way. Because our analysis shows the sensitivity of
$C'$ to the NLO corrections, it is worthwhile to investigate the
direct CP asymmetries of other charged $B$ meson decays. The
results will be published elsewhere. The color-suppressed tree
amplitude $C$ involved in the $B\to\pi\pi$ decays, despite of
being increased few times too by the vertex corrections, remains
subleading with the ratio $|C/T|\approx 0.2$, where $T$ represents
the color-allowed tree amplitude. This ratio is not enough to
explain the observed $B^0\to\pi^0\pi^0$ branching ratio as shown
in Table~\ref{br1} \cite{LMS05}. A much larger $|C/T|\approx 0.8$
must be achieved in order to resolve the $B\to\pi\pi$ puzzle
\cite{Charng2}. We mention that a different source for the large
relative strong phase between $C$ and $T$ has been proposed in
\cite{GHZP}, which arises from charm- and top-mediated penguins.

\subsection{Helicity Amplitudes}

We examine whether the observations made in \cite{LMS05} are solid
by applying the same NLO PQCD formalism to the $B\to\rho\rho$
decays, which are also sensitive to the color-suppressed tree
contribution. The $B\to\rho\rho$ decays have been analyzed at LO
in \cite{LILU,rho}. The numerical results in the two references
differ a bit due to the different choices of the characteristic
hard scales, which can be considered as one of the sources of
theoretical uncertainties (from higher-order corrections). The
$B\to V_2(\epsilon_2,P_2) V_3(\epsilon_3,P_3)$ decay rate is
written as
\begin{equation}
\Gamma =\frac{G_{F}^{2}P_c}{64\pi m^{2}_{B} } \sum_{\sigma}{\cal
M}^{(\sigma)\dagger }{\cal M^{(\sigma)}}\;, \label{dr1}
\end{equation}
where $P_c=|{\bf P}_2|=|{\bf P}_3|=m_B/2$ is the momentum of
either of the vector mesons $V_2$ and $V_3$, $m_B$ being the $B$
meson mass. $\epsilon_{2}$ $(\epsilon_{3})$ are the polarization
vectors of the meson $V_2$ $(V_3)$. The amplitudes $\cal
M^{(\sigma)}$ corresponding to the polarization configurations
$\sigma$ with both $V_2$ and $V_3$ being longitudinally polarized,
and being transversely polarized in the parallel and perpendicular
directions are written as
\begin{eqnarray}
{\cal M}^{\sigma} \ =\ \left( m_{B}^{2}{\cal M}_{L}\;,\ \
m_{B}^{2}{\cal M}_{N}
\epsilon^{*}_{2}(T)\cdot\epsilon^{*}_{3}(T)\;,\ \ - i{\cal
M}_{T}\epsilon^{\alpha \beta\gamma \rho}
\epsilon^{*}_{2\alpha}(T)\epsilon^{*}_{3\beta}(T) P_{2\gamma
}P_{3\rho } \right) \;,
\end{eqnarray}
respectively. In the above expressions $\epsilon(T)$ denote the
transverse polarization vectors, and we have adopted the
convention $\epsilon_{0123} = 1$.

Define the velocity $v_2=P_2/m_{V_2}$ $(v_3=P_3/m_{V_3})$ in terms
of the $V_2$ $(V_3)$ meson mass $m_{V_2}$ $(m_{V_3})$. The
helicity amplitudes,
\begin{eqnarray}
A_{L}&=&-G m^{2}_{B}{\cal M}_{L}, \nonumber\\
A_{\parallel}&=&G \sqrt{2}m^{2}_{B}{\cal M}_{N}, \nonumber \\
A_{\perp} &=&
G m_{V_2} m_{V_3} \sqrt{2[(v_2\cdot v_3)^{2}-1]} {\cal M }_{T}
\;, \label{ase3}
\end{eqnarray}
with the normalization factor $G=\sqrt{G_F^2P_c/(64\pi
m^2_{B}\Gamma)}$, satisfy the relation,
\begin{eqnarray}
|A_{L}|^2+|A_{\parallel}|^2+|A_{\perp}|^2=1\;.
\end{eqnarray}
We also need to employ another equivalent set of helicity
amplitudes,
\begin{eqnarray}
H_{0}\ =\ m^{2}_{B} {\cal M}_L\;, \ \ \ \ \ \ \ H_{\pm}\ =\
m^{2}_{B} \left( {\cal M}_{N} \mp  \frac{{\cal M}_{T}}{2} \right)
\;,\label{ase2}
\end{eqnarray}
with the helicity summation,
\begin{eqnarray}
\sum_{\sigma}{\cal M}^{(\sigma)\dagger }{\cal M^{(\sigma)}} =
|H_{0}| ^{2}+|H_{+}|^{2} + | H_{-}|^{2}\;.
\end{eqnarray}
The definitions in Eq.~(\ref{ase3}) are related to those in
Eq.~(\ref{ase2}) via
\begin{eqnarray}
A_{L}\ =\ -G H_0 \;,\ \ \ \ \ A_{\parallel}\ =\
\frac{G}{\sqrt{2}}(H_++H_-) \;,\ \ \ \ \ A_{\perp}\ =\
-\frac{G}{\sqrt{2}}(H_+-H_-) \;. \label{ase}
\end{eqnarray}

The explicit expressions of the distribution amplitudes
$\phi_\rho$, $\phi_{\rho}^t$, and $\phi_{\rho}^s$ for a
longitudinally polarized $\rho$ meson, and $\phi_\rho^T$,
$\phi_{\rho}^v$, and $\phi_{\rho}^a$ for a transversely polarized
$\rho$ meson are referred to \cite{TLS,PB1}. However, for the
twist-3 distribution amplitudes $\phi_{\rho}^t$, $\phi_{\rho}^s$,
$\phi_{\rho}^v$, and $\phi_{\rho}^a$, we adopt their asymptotic
models as shown below:
\begin{eqnarray}
\phi_\rho(x)&=&\frac{3f_\rho}{\sqrt{2N_c}} x(1-x)\left[1+
0.18\, C_2^{3/2}(2x-1)\right]\;,
\label{pwr}\\
\phi_{\rho}^t(x)&=&\frac{f^T_{\rho}}{2\sqrt{2N_c}}
3(2x-1)^2
\;,
\label{pwt}\\
\phi_{\rho}^s(x) &=&\frac{3f_\rho^T}{2\sqrt{2N_{c}}}
(1-2x)
\;,
\label{pws}\\
\phi_\rho^T(x)&=&\frac{3f_\rho^T}{\sqrt{2N_c}} x(1-x)\left[1+
0.2\, C_2^{3/2}(2x-1)\right]\;,
\label{pwft}\\
\phi_{\rho}^v(x)&=&\frac{f_{\rho}}{2\sqrt{2N_c}}
\frac{3}{4}[1+(2x-1)^2]
\;,
\label{pwv}\\
\phi_{\rho}^a(x) &=&\frac{3f_\rho}{4\sqrt{2N_{c}}}
(1-2x)
\;, \label{pwa}
\end{eqnarray}
with the decay constants $f_\rho=200$ MeV and $f_\rho^T=160$ MeV,
and the Gegenbauer polynomial $C_2^{3/2}(t)=3(5 t^2-1)/2$. On one
hand, the sum-rule derivation of light-cone meson distribution
amplitudes suffer sizable theoretical uncertainty, so that the
asymptotic models are acceptable. On the other hand, the
asymptotic models for twist-3 distribution amplitudes were also
adopted in QCDF \cite{BBNS}, and the comparison of our results
with theirs will be more consistent.

\begin{table}[hb]
\begin{center}
\begin{tabular}{l|c}
\hline \hline $\rho^+\rho^-$& $H_{h}^{(u)} $
\\\hline
$F_{e}^h$ & $F^h_{e4} \left( a_1 \right)$
\\
${\cal M}_{e}^h$ & ${\cal M}^h_{e4} \left( a_1' \right)$
\\
$F_{a}^h$ & $\eta_T  F^h_{a4} \left( a_2 \right)$
\\
${\cal M}_{a}^h$ & ${\cal M}^h_{a4} \left( a_2' \right)$
\\\hline
$\rho^+\rho^-$& $H_{h}^{(t)} $
\\\hline
$F_e^{P,h}$ & $F^h_{e4} \left( a_4^{(u)} \right) $
\\
${\cal M}_{e}^{P,h}$ & ${\cal M}^h_{e4}\left( a_4^{\prime (u)}
\right) + {\cal M}^h_{e6}\left( a_6^{\prime (u)} \right)$
\\
$F_a^{P,h}$ & $ \eta_T  F^h_{a4} \left( a_4^{(d)} \right) +
            F^h_{a6} \left(a_6^{(d)} \right)$
\\
${\cal M}_{a}^{P,h}$ & ${\cal M}^h_{a4}
  \left(  a_3^{\prime (u)} + a_3^{\prime (d)} +  a_4^{\prime (d)}
 + a_5^{\prime (u)} + a_5^{\prime (d)}\right)
+ \eta_T {\cal M}^h_{a6} \left( a_6^{\prime (d)} \right)$
\\\hline $\rho^+\rho^0$  & $\sqrt{2}H_{h}^{(u)} $
\\\hline
$F^h_e$ & $F^h_{e4} \left( a_1 + a_2 \right)$
\\
${\cal M}^h_{e}$ & ${\cal M}^h_{e4} \left( a_1' + a_2' \right)$
\\
$F^h_a$ & 0
\\
${\cal M}^h_{a}$ & 0
\\\hline
$\rho^+\rho^0$ & $\sqrt{2}H_{h}^{(t)} $
\\\hline
$F_e^{P,h}$ & $F^h_{e4}
  \left( a_3^{(u)} - a_3^{(d)} + a_4^{(u)} - a_4^{(d)}
         + a_5^{(u)} - a_5^{(d)} \right)$
\\
${\cal M}_{e}^{P,h}$ & ${\cal M}^h_{e4}\left( a_3^{\prime (u)} -
a_3^{\prime (d)}
                       + a_4^{\prime (u)} - a_4^{\prime (d)}
- a_5^{\prime (u)} + a_5^{\prime (d)} \right) + {\cal
M}^h_{e6}\left( a_6^{\prime (u)} - a_6^{\prime (d)}\right) $
\\
$F_a^{P,h}$   & 0
\\
${\cal M}_{a}^{P,h}$   & 0
\\\hline $\rho^0\rho^0$ & $\sqrt{2}H_{h}^{(u)} $
\\\hline
$F^h_e$ & $F^h_{e4} \left(- a_2 \right)$
\\
${\cal M}^h_{e}$ & ${\cal M}^h_{e4} \left( -a_2' \right)$
\\
$F^h_a$ & $\eta_T  F^h_{a4} \left( a_2 \right)$
\\
${\cal M}^h_{a}$ & ${\cal M}^h_{a4} \left( a_2' \right)$
\\\hline
$\rho^0\rho^0$ & $\sqrt{2}H_{h}^{(t)} $
\\\hline
$F_e^{P,h}$ & $ F^h_{e4} \left( -a_3^{(u)} + a_3^{(d)} + a_4^{(d)}
- a_5^{(u)} + a_5^{(d)} \right)$
\\
${\cal M}_{e}^{P,h}$ & ${\cal M}^h_{e4}\left( - a_3^{\prime (u)} +
a_3^{\prime (d)}
                     + a_4^{\prime (d)}
+ a_5^{\prime (u)} - a_5^{\prime (d)} \right) + {\cal
M}^h_{e6}\left( a_6^{\prime (d)} \right)$
\\
$F_a^{P,h}$ & $ \eta_T  F^h_{a4} \left( a_4^{(d)} \right) +
            F^h_{a6} \left(a_6^{(d)} \right)$
\\
${\cal M}_{a}^{P,h}$ & ${\cal M}^h_{a4} \left( a_3^{\prime (u)} +
a_3^{\prime (d)}
                       +  a_4^{\prime (d)}
 + a_5^{\prime (u)} + a_5^{\prime (d)}\right)
+ \eta_T {\cal M}^h_{a6} \left( a_6^{\prime (d)} \right)$
\\\hline\hline
\end{tabular}
\caption{LO $B\to \rho\rho$ decay amplitudes with $\eta_{T}=0\
(1)$ for the longitudinal (transverse) components.} \label{rhorho}
\end{center}
\end{table}

For the $\bar b \to \bar d$ transition, the helicity amplitudes
have the general expression,
\begin{eqnarray}
H_{h} &=& V_{ub}^*V_{ud}\, H_{h}^{(u)}
 +V_{cb}^*V_{cd}\, H_{h}^{(c)}
 +V_{tb}^*V_{td}\, H_{h}^{(t)}
\;, \label{eq:amp}
\end{eqnarray}
with $h=0$ or $\pm$, and $V$'s being the Cabibbo-Kobayashi-Maskawa
(CKM) matrix elements. The amplitudes $H^{(u)}_{h}$,
$H^{(c)}_{h}$, and $H^{(t)}_{h}$ are decomposed at LO into
\begin{eqnarray}
H^{(u)}_{h} &=& m_B^2 \left( f_\rho F^h_e + {\cal M}^h_e +
f_BF^h_a + {\cal M}^h_a \right) \;,
\nonumber\\
H^{(c)}_{h} &=& 0 \;,
\nonumber\\
H^{(t)}_{h} &=& - m_B^2 \left( f_\rho F^{P,h}_e + {\cal M}^{P,h}_e
+ f_BF^{P,h}_a + {\cal M}^{P,h}_a \right) \;.
\end{eqnarray}
The LO PQCD factorization formulas for the $B\to\rho\rho$ helicity
amplitudes associated with the final states $\rho^+\rho^-$,
$\rho^+\rho^0$, and $\rho^0\rho^0$ are summarized in
Table~\ref{rhorho}. The Wilson coefficients $a^{(q)}$ for the
factorizable contributions, and $a^{\prime(q)}$ for the
nonfactorizable contributions can be found in \cite{LMS05}, where
$q=u$ or $d$ denotes the quark pair produced in the electroweak
penguin.

The explicit expressions of the LO factorizable amplitudes
$F^0_{e4,a4,a6}$ and of the LO nonfactorizable amplitudes ${\cal
M}^0_{e4,e6,a4,a6}$ are similar to those for the $B\to PP$ decays
\cite{LMS05} but with the replacements of the distribution
amplitudes and the masses,
\begin{eqnarray}
& &\phi^A(x)\to\phi(x)\;,\;\;\;\;\phi^P(x)\to\phi^s(x)\;,\;\;\;\;
\phi^T(x)\to\phi^t(x)\;,\nonumber\\
& &m_{02}\to -m_\rho\;,\;\;\;\;m_{03}\to m_\rho\;. \label{REPL}
\end{eqnarray}
In the above replacement $m_{02}$ ($m_{03}$) is the chiral
enhancement scale associated with the pseudo-scalar meson involved
in the $B\to P$ transition (emitted from the weak vertex), and
$m_\rho=0.77$ GeV the $\rho$ meson mass. Note that the amplitude
$F^0_{e6}$ from the operators $O_{5-8}$ vanishes at LO. The LO
factorization formulas for the transverse components are collected
in Appendix A, whose relations to $F^\pm$ and to ${\cal M}^\pm$ in
Table~\ref{rhorho} follow Eq.~(\ref{ase2}). For example, the
amplitude $F^\pm_{e4}$ is given by
\begin{eqnarray}
F^\pm_{e4}\ =\ F_{Ne4} \mp \frac{F_{Te4}}{2} \;.
\end{eqnarray}

\subsection{NLO Corrections}

The vertex corrections to the $B\to \rho\rho$ decays modify the
Wilson coefficients for the emission amplitudes in the standard
definitions \cite{LMS05} into
\begin{eqnarray}
a_1(\mu) &\to& a_1(\mu)
+\frac{\alpha_s(\mu)}{4\pi}C_F\frac{C_{1}(\mu)}{N_c} V_1(\rho) \;,
\nonumber\\
a_2(\mu) &\to& a_2(\mu)
+\frac{\alpha_s(\mu)}{4\pi}C_F\frac{C_{2}(\mu)}{N_c} V_2(\rho) \;,
\nonumber\\
a_i(\mu) &\to& a_i(\mu)
+\frac{\alpha_s(\mu)}{4\pi}C_F\frac{C_{i\pm 1}(\mu)}{N_c}
V_i(\rho) \;,\;\;\;\;i=3 - 10\;,\label{wnlo}
\end{eqnarray}
where $V_i(\rho)$ in the NDR scheme are in agreement with those in
\cite{BN} for the longitudinal component,
\begin{eqnarray}
V_i(\rho) &=& \left\{ {\renewcommand\arraystretch{2.5}
\begin{array}{ll}
12\ln\displaystyle{\frac{m_b}{\mu}}-18
+\frac{2\sqrt{2N_c}}{f_\rho}\int_0^1 dx\, \phi_\rho(x)\, g(x)\;, &
\mbox{\rm for }i=1-4,9,10\;,
\\
-12\ln\displaystyle{\frac{m_b}{\mu}}+6
-\frac{2\sqrt{2N_c}}{f_\rho}\int_0^1dx\, \phi_\rho(x)\, g(1-x)\;,
& \mbox{\rm for }i=5,7\;,
\\
\displaystyle{ -\frac{2\sqrt{2N_c}}{f_\rho^T}\int_0^1 dx\,
\phi_\rho^{s}(x)\, [-6+h(x)] }\;, & \mbox{\rm for }i=6,8\;,
\end{array}
} \right.\label{vim}
\end{eqnarray}
and with those in \cite{YWL} for the transverse components,
\begin{eqnarray}
V_i^\pm(\rho) &=& \left\{ {\renewcommand\arraystretch{2.5}
\begin{array}{ll}
12\ln\displaystyle{\frac{m_b}{\mu}}-18
+\frac{2\sqrt{2N_c}}{f_\rho}\int_0^1 dx\,
[\phi_\rho^v(x)\pm\phi_\rho^a(x)]\, g(x)\;, & \mbox{\rm for
}i=1-4,9,10\;,
\\
-12\ln\displaystyle{\frac{m_b}{\mu}}+6
-\frac{2\sqrt{2N_c}}{f_\rho}\int_0^1dx\,
[\phi_\rho^v(x)\pm\phi_\rho^a(x)]\, g(1-x)\;, & \mbox{\rm for
}i=5,7\;.
\end{array}
} \right.\label{vit}
\end{eqnarray}
We do not show $V_{6,8}^\pm$, because of the associated
factorizable emission amplitudes $F^{\pm}_{e6}=0$. Moreover, the
vertex corrections introduce the additional contributions
resulting from the penguin operators $O_{5-8}$,
\begin{eqnarray}
&\rho^+\rho^-: &f_\rho F^{P,h}_e\to f_\rho F^{P,h}_e+f_\rho^T
F^h_{e6} \left( a_{6\rm VC}^{(u)} \right)\;,\nonumber\\
&\rho^+\rho^0: &f_\rho F^{P,h}_e\to f_\rho F^{P,h}_e+f_\rho^T
F^h_{e6} \left( a_{6\rm VC}^{(u)}-a_{6\rm VC}^{(d)} \right)\;,\nonumber\\
&\rho^0\rho^0: &f_\rho F^{P,h}_e\to f_\rho F^{P,h}_e+f_\rho^T
F^h_{e6} \left( a_{6\rm VC}^{(d)} \right)\;,
\end{eqnarray}
where the arguments $a_{6\rm VC}$ represent only the
vertex-correction piece in Eq.~(\ref{wnlo}).

Taking into account the NLO contributions from the quark loops and
from the magnetic penguin, the helicity amplitudes are modified
into
\begin{eqnarray}
{\renewcommand\arraystretch{2.0}
\begin{array}{lll}
\displaystyle \rho^+\rho^-: &H^{(u,c)}_{h} \, \to\, H^{(u,c)}_{h}
+m_B^2{\cal M}^{(u,c)}_{h}\;, & \displaystyle H^{(t)}_{h} \, \to\,
H^{(t)}_{h}
 - m_B^2 {\cal M}^{(t)}_{h}
 - m_B^2{\cal M}^{(g)}_{h}\;,
\\
\displaystyle \rho^+\rho^0:& H^{(u,c,t)}_{h} \, \to\,
H^{(u,c,t)}_{h}\;, &
\\
\displaystyle \rho^0\rho^0:&
\displaystyle
 H^{(u,c)}_{h} \, \to\, H^{(u,c)}_{h}
  + \frac{m_B^2}{\sqrt{2}}{\cal M}^{(u,c)}_{h}\;,
&\displaystyle
 H^{(t)}_{h} \, \to\,
  H^{(t)}_{h}
-\frac{m_B^2}{\sqrt{2}}{\cal M}^{(t)}_{h} -\frac{m_B^2}{\sqrt{2}}{\cal
M}^{(g)}_{h} \;,
\end{array}
}\label{qlmp}
\end{eqnarray}
where ${\cal M}_{h}^{(u)}$, ${\cal M}_{h}^{(c)}$, ${\cal
M}_{h}^{(t)}$, and ${\cal M}_{h}^{(g)}$ denote the up-loop,
charm-loop, QCD-penguin-loop, and magnetic-penguin corrections,
respectively. The magnetic-penguin contribution to the $B\to PV$
modes was computed in \cite{MISHIMA03}. ${\cal M}_{h}^{(u,c,t)}$
and ${\cal M}_{h}^{(g)}$ for $h=0$ are similar to those for the
$B\to PP$ decays \cite{LMS05} with the replacements in
Eq.~(\ref{REPL}). Those for the transverse components are
presented in Appendix A.

The choices of the $B$ meson wave function, of the $B$ meson
lifetimes, and of the CKM matrix elements, including the allowed
ranges of their variations, are the same as in \cite{LMS05}. We
vary the Gegenbauer coefficients in $\phi_\rho$ and in
$\phi_\rho^T$ by 100\% as analyzing the theoretical uncertainty.
The resultant $B\to\rho$ form factors at maximal recoil,
\begin{eqnarray}
A_0\ =\ 0.31^{+0.07}_{-0.06} \;,\;\;\;\; A_1\ =\
0.21^{+0.05}_{-0.04} \;,\;\;\;\; V\ =\ 0.26^{+0.07}_{-0.05}
\;,\label{form}
\end{eqnarray}
associated with the longitudinal, parallel, and perpendicular
components of the $B\to\rho\rho$ decays, respectively, are similar
to those derived from QCD sum rules \cite{sumrho,BZ0412}, and
almost the same as adopted in the QCDF analysis \cite{AK}.
Compared to \cite{sumrho}, one-loop radiative corrections to the
two-parton twist-3 contributions have been considered in
\cite{BZ0412}. The central value of the form factor $V$ in
Eq.~(\ref{form}) is a bit smaller than those in
\cite{sumrho,BZ0412}. We emphasize that this difference is not
essential, since the perpendicular component corresponding to $V$
contributes roughly less than 10\% of the total $B\to\rho\rho$
branching ratios as shown below.

The PQCD results for the $B\to\rho\rho$ branching ratios, together
with the BABAR and Belle data, are listed in Table~\ref{br2}. It
is obvious that the NLO PQCD values are consistent with the data
of the $B^0 \to \rho^\mp \rho^\pm$ and $B^\pm \to \rho^\pm \rho^0$
branching ratios. The color-suppressed tree amplitude is also
enhanced by the vertex corrections here, but the ratio
$|C/T|\approx 0.2$ for the longitudinal component, similar to that
in the $B\to\pi\pi$ decays, is still small. However, the central
value of the predicted $B^0 \to \rho^0 \rho^0$ branching ratio has
almost saturated the experimental upper bound. We conclude that it
is unlikely to accommodate the measured $B^0\to\pi^0\pi^0$,
$\rho^0\rho^0$ branching ratios simultaneously in PQCD.

\begin{table}[hbt]
\begin{center}
\begin{tabular}{cccccccccc}
\hline\hline Mode & BABAR \cite{HFAG}  & Belle \cite{HFAG} & LO &
LO$_{\rm NLOWC}$ & +VC & +QL &  +MP  & +NLO
\\
\hline $B^0 \to \rho^\mp \rho^\pm$ & $ 30 \pm 4\pm 5 $ & $22.8\pm
3.8^{+2.3}_{-2.6}$ & $27.8$& $26.1$&$25.2$&$26.6$&$25.9$&
$25.3^{+25.3\,(+12.1)}_{-13.8\,(-\ 7.9)}$
\\
$B^\pm \to \rho^\pm \rho^0$ & $17.2\pm 2.5\pm 2.8$ & $31.7\pm
7.1^{+3.8}_{-6.7}$ & $13.7$& $16.2$&$16.0$&$16.2$&$16.2$&
$16.0^{+15.0\,(+\ 7.8)}_{-\ 8.1\,(-\ 5.3)}$
\\
$B^0 \to \rho^0 \rho^0$ & $ <1.1 $ & --- &
$0.33$&$0.56$&$1.02$&$0.62$&$0.45$&
$0.92^{+1.10\,(+0.64)}_{-0.56\,(-0.40)}$
\\
\hline\hline
\end{tabular}
\end{center}
\caption{$B\to\rho\rho$ branching ratios from PQCD in the NDR
scheme in units of $10^{-6}$.}\label{br2}
\end{table}

We obtain the direct CP asymmetries $A_{CP}(B^0 \to \rho^\mp
\rho^\pm)=-0.02\, (-0.07)$, $A_{CP}(B^\pm \to
\rho^\pm\rho^0)=0.00\, (0.00)$, and $A_{CP}(B^0 \to \rho^0
\rho^0)=0.56\, (0.80)$, where the values (in the parentheses) are
from LO (NLO) PQCD. We have also computed the polarization
fractions. The NLO corrections have a minor impact on the $B^0 \to
\rho^\mp \rho^\pm$ and $B^\pm \to \rho^\pm \rho^0$ decays: their
longitudinal polarization contributions remain dominant, reaching
93\% and 97\%, respectively. However, the ${\bar B}^0 \to \rho^0
\rho^0$ polarization fractions are sensitive to the NLO
corrections as indicated in Table~\ref{pof}, where the average
longitudinal, parallel, and perpendicular polarization fractions,
$f_L$, $f_\parallel$, and $f_\perp$, respectively, are defined by
\begin{eqnarray}
f_{L,\parallel,\perp}=\frac{B(B^0 \to \rho^0
\rho^0)_{L,\parallel,\perp}+B({\bar B}^0 \to \rho^0
\rho^0)_{L,\parallel,\perp}}{B(B^0 \to \rho^0 \rho^0)+B({\bar B}^0
\to \rho^0 \rho^0)}\;.
\end{eqnarray}
The average longitudinal polarization fraction of the $B^0 \to
\rho^0 \rho^0$ decays was also found to be smaller in LO PQCD
\cite{LILU,rho}. It is easy to understand the changes due to the
NLO effects. As stated before, the color-suppressed tree
amplitude, being the main tree contribution in the $B^0 \to \rho^0
\rho^0$ decay, is enhanced by the vertex corrections. The $B^0 \to
\rho^0 \rho^0$ polarization fractions should then approach the
naive counting rules \cite{CKL2,AK,LM04}: $f_L\sim 1$ and
$f_\parallel\sim f_\perp \sim \lambda^2$ obeyed by a
tree-dominated decay, where $\lambda\approx 0.22$ is the
Wolfenstein parameter.

\begin{table}[hbt]
\begin{center}
\begin{tabular}{cccc}
\hline\hline Mode & $f_L$ & $f_\parallel$ & $f_\perp$
\\
\hline $B^0 \to \rho^0 \rho^0$ & 0.71 (0.67) &0.14 (0.15) & 0.15
(0.18)
\\
${\bar B}^0 \to \rho^0 \rho^0$ & 0.09 (0.79) & 0.45 (0.10) &0.46
(0.11)
\\
Average & 0.23 (0.78) & 0.38 (0.11)& 0.39 (0.11)
\\
\hline\hline
\end{tabular}
\end{center}
\caption{LO and NLO (in the parentheses) polarization fractions of
the $B^0\to\rho^0\rho^0$ decays from PQCD.}\label{pof}
\end{table}

\section{JET FUNCTION IN SCET}

In this section we investigate the resolution to the $B\to\pi\pi$
puzzle claimed in QCDF with the input of the NLO jet function from
SCET \cite{BY05}. The leading-power SCET formalism has been
derived for two-body nonleptonic $B$ meson decays \cite{BPRS}.
However, there exist different opinions on the calculability of
the hard coefficients in SCET$_{\rm II}$, one of which is the jet
function characterized by a scale of $O(\sqrt{m_b\Lambda})$. In
\cite{BPS05} the jet function is regarded as being incalculable,
and treated as a free parameter. Together with other hadronic
parameters, it is determined by fitting the SCET formalism to the
$B\to\pi\pi$ data. Therefore, the large ratio $|C/T|$ obtained in
\cite{BPS05} is an indication of the data, instead of coming from
an explicit evaluation of the amplitudes. In this analysis the QCD
penguin amplitude, receiving a significant contribution from the
long-distance charming penguin \cite{CNPR}, was also found to be
important. Similarly, the large charming penguin, as one of the
fitting parameters in SCET, also arises from the data fitting. A
global analysis of the $B\to\pi\pi$, $\pi K$ decays based on the
leading-power SCET parametrization has been performed recently in
\cite{WZ0610}, where a smaller branching ratio
$B(B^0\to\pi^0\pi^0)\approx 0.84\times 10^{-6}$ was obtained.

A plausible mechanism in SCET for enhancing the ratio $|C/T|$ was
provided in \cite{BY05}: the jet function could increase the
nonfactorizable spectator contribution to the color-suppressed
tree amplitude $C$ at NLO. This significant effect was implemented
into QCDF \cite{BY05}. Because of the end-point singularities
present in twist-3 spectator amplitudes and in annihilation
amplitudes, these contributions have to be parameterized in QCDF
\cite{BBNS}. Different scenarios for choosing the free parameters,
labelled by ``default", ``S1", ``S2", $\cdots$, ``S4", were
proposed in \cite{BN}. As shown in Table~\ref{tab1}, the large
measured $B^0\to\pi^0\pi^0$ branching ratio can be accommodated,
when the parameter scenario S4 is adopted. It has been emphasized
in the Introduction that the same formalism should be applied to
other decay modes for a check, among which we focus on the
quantities sensitive to $C$: the $B\to \pi K$ direct CP
asymmetries and the $B^0\to\rho^0\rho^0$ branching ratio.

The QCDF formulas for the $B\to VV$ decays with the NLO
contributions from the vertex corrections, the quark loops, and
the magnetic penguin can be found in \cite{AK,CY01,YWL}, which
appear as the $O(\alpha_s)$ terms of the Wilson coefficients
$a_i$, $i=1,\cdots,10$. The vertex corrections are the same as in
Eqs.~(\ref{vim}) and (\ref{vit}). Note that the expressions of the
Wilson coefficients $a_{6,8}$ differ between \cite{AK} and
\cite{CY01,YWL}: $a_{6,8}$ for both the longitudinal and
transverse components in \cite{CY01,YWL} do not receive any
$O(\alpha_s)$ correction. We disagree on this result as shown in
Eqs.~(\ref{vim}) and (\ref{qlmp}). Hence, we adopt the expressions
in \cite{AK} for the contributions from the quark loops, the
magnetic penguin, and the annihilation. We also employ the
$B\to\rho$ form factor values in \cite{AK}. Since the spectator
amplitudes were not shown explicitly in \cite{AK}, we use those
from \cite{BN}. The parameter sets default and S4 have been
defined for the $B\to PP$ decays \cite{BN}, but have not for the
$B\to VV$ ones. Therefore, we assume that the parameters for the
latter are the same as for the former in the following analysis.
Fortunately, the predicted $B^0\to\rho^0\rho^0$ branching ratio is
insensitive to the variation of the annihilation phase $\phi_A$,
which is one of the most essential parameters in QCDF: varying
$\phi_A$ between 0 and $2\pi$, the $B^0\to\rho^0\rho^0$ branching
ratio changes by less than 10\%.

The jet function $j_\parallel$ derived in \cite{BY05} is relevant
to the $B\to PP$ decays and to the $B\to VV$ decays with
longitudinally polarized final states. The jet function $j_\perp$
is relevant to the $B\to VV$ decays with transversely polarized
final states. These jet functions apply not only to the
color-suppressed tree amplitudes, but to the color-allowed tree
and penguin amplitudes, which are free of the end-point
singularities. We mention that the NLO corrections to the hard
coefficients of SCET$_{I}$ have been derived in \cite{BJ05,BJ052}.
This new piece modifies the QCDF outcomes slightly, comparing the
color-allowed and color-suppressed tree contributions obtained in
\cite{BY05} and in \cite{BJ05}. Hence, we consider the NLO
correction only from the jet function for simplicity. Furthermore,
since the jet function enhances the color-suppressed tree
amplitude, the $B^0\to \rho^0\rho^0$ polarization fractions are
expected to approach the naive counting rules. That is, the
longitudinal component dominates. This tendency has been confirmed
in PQCD as indicated by Table~\ref{pof}. To serve our purpose, it
is enough to evaluate only the $B\to \rho_L\rho_L$ branching
ratios here.

\begin{table}[hbt]
\begin{center}
\begin{tabular}{cccccc}
\hline\hline Mode&Data \cite{HFAG} & default, LO jet  &
default, NLO jet & S4, LO jet & S4, NLO jet
\\\hline
$B^\pm\to\pi^\pm\pi^0$& $ \phantom{0}5.5 \pm 0.6 $ &
\phantom{0}6.02 (6.03) & \phantom{0}6.24 (6.28) &
\phantom{0}5.07 (5.07) & \phantom{0}5.77 (5.87)
\\
$B^0\to\pi^\mp\pi^\pm$& $ \phantom{0}4.9 \pm 0.4 $ &
\phantom{0}8.90 (8.86) & \phantom{0}8.69 (8.62) & \phantom{0}5.22
(5.17) & \phantom{0}4.68 (4.58)
\\
$B^0\to\pi^0\pi^0$& $ \phantom{0}1.45 \pm 0.29 $ &
\phantom{0}0.36 (0.35) & \phantom{0}0.40 (0.40) &
\phantom{0}0.72 (0.70)  & \phantom{0}1.07 (1.13)
\\
$B^\pm \to \pi^\pm K^0$&$ 24.1 \pm 1.3 $ &
20.50 (19.3) &20.13 &21.60 (20.3)&20.50
\\
$B^\pm \to \pi^0 K^\pm$& $ 12.1 \pm 0.8 $&
11.79 (11.1)  &11.64 &12.48 (11.7)&12.02
\\
$B^0 \to \pi^\mp K^\pm$&$ 18.9 \pm 0.7 $&
17.33 (16.3) &17.21 &19.60 (18.4)&19.23
\\
$B^0\to\pi^0 K^0$&$ 11.5 \pm 1.0 $&
\phantom{0}7.49 (\phantom{0}7.0) &\phantom{0}7.41 &
\phantom{0}8.56 (\phantom{0}8.0)&\phantom{0}8.36
\\
$B^\pm\to\rho^\pm_L\rho^0_L$&$19.1\pm 3.5$ & 18.51 &19.48 &16.61 &
18.64
\\
$B^0\to\rho^\mp_L\rho^\pm_L$&$25.2^{+3.6}_{-3.7}$ & 25.36 & 24.42
& 18.48 & 16.76
\\
$B^0\to\rho^0_L\rho^0_L$&$<1.1$ &
\phantom{0}0.43&\phantom{0}0.66&\phantom{0}0.92&\phantom{0}1.73
\\
\hline\hline
\end{tabular}
\caption{Branching ratios from QCDF with the input of the SCET jet
function in units of $10^{-6}$. The values in the parentheses are
quoted from \cite{BY05,BN} for comparison. The data for the
$B\to\rho\rho$ decays include all polarizations.} \label{tab1}
\end{center}
\end{table}

\begin{table}[hbt]
\begin{center}
\begin{tabular}{cccccc}
\hline\hline Mode&Data \cite{HFAG} & default, LO jet  & default,
NLO jet & S4, LO jet & S4, NLO jet
\\\hline
$B^\pm\to\pi^\pm\pi^0$& $ \phantom{-}1 \pm 6 $ &
$\ \, -0.02$ ($-0.02$) &$-0.02$ &$\ \, -0.02$ ($-0.02$)  &$\ \, -0.02$
\\
$B^0\to\pi^\mp\pi^\pm$& $ \phantom{-}37 \pm 10$ &
$-6.57$ ($-6.5$) &$-6.65$ &$\phantom{-}10.60$ ($\phantom{-}10.3$) &
$\phantom{-}10.91$
\\
$B^0\to\pi^0\pi^0$& $ \phantom{-}28^{+40}_{-39}$ &
$\ 44.67$ ($\ 45.1$) &
$\ 41.95$ & $-19.58$ ($-19.0$) &$-18.48$
\\
$B^\pm\to \pi^\pm K^0$& $ -2 \pm  4$  &
$\phantom{-}0.84$ ($\phantom{-}0.9$) & $\phantom{-}0.85$ &
$\phantom{-}0.29$ ($\phantom{-}0.3$)&$\ \,\phantom{-}0.29$
\\
$B^\pm\to \pi^0 K^\pm$ & $ \phantom{-}4 \pm 4 $ &
$\phantom{-}6.88$ ($\phantom{-}7.1$) &$\phantom{-}7.04$ &
$-3.53$ ($-3.6$)  &$\ \,-4.08$
\\
$B^0\to \pi^\mp K^\pm$& $ -10.8 \pm 1.7 $ & $\phantom{-}4.28$
($\phantom{-}4.5$) & $\phantom{-}4.24$& $-4.06$ ($-4.1$)&$\
\,-3.89$
\\
$B^0\to \pi^0 K^0$& $ \phantom{-}2 \pm 13 $ &
$-3.15$ ($-3.3$) & $-3.37$ &$\phantom{-}0.78$ ($\phantom{-}0.8$)&
$\ \,\phantom{-}1.60$
\\
\hline\hline
\end{tabular}
\caption{Direct CP asymmetries from QCDF with the input of the
SCET jet function in percentage. The values in the parentheses are
quoted from \cite{BN} for comparison.} \label{tab2}
\end{center}
\end{table}

The predictions for the $B\to\pi\pi$, $\pi K$, $\rho_L\rho_L$
decays from QCDF with the input of the SCET jet function are
summarized in Tables~\ref{tab1} and \ref{tab2}. The values in the
parentheses are quoted from \cite{BY05} for the $B\to\pi\pi$
branching ratios, and from \cite{BN} for the $B\to\pi\pi$ direct
CP asymmetries and for the $B\to\pi K$ decays. The small
differences between our results and those from \cite{BY05,BN} are
attributed to the different choices of the CKM matrix elements,
meson masses, etc. All the calculations performed in this work,
except those of the $B\to\pi\pi$ branching ratios, are new. It is
found that the scenario S4 plus the NLO jet function lead to the
ratio $C/T\approx 0.8$, and accommodate at least the BABAR data of
the $B^0\to\pi^0\pi^0$ branching ratio. Nevertheless, the same
configuration overshoots the experimental upper bound of the
$B^0\to\rho^0\rho^0$ branching ratio apparently, implying that the
color-suppressed tree amplitude is enhanced overmuch by the NLO
jet function. Adopting the default scenario, QCDF satisfies the
$B^0\to\rho^0\rho^0$ bound, but the predicted $B^0\to\pi^0\pi^0$
branching ratio becomes too small. We have surveyed the other
scenarios, and found the results from S1 and S3 (S2) similar to
those from the default (S4). That is, it is also unlikely to
accommodate the $B\to\pi\pi$, $\rho\rho$ data simultaneously in
QCDF. The $B\to\pi K$ branching ratios are not affected by the NLO
jet function, because the color-suppressed tree amplitude is still
subleading in the penguin-dominated modes. The
$B^\pm\to\rho^\pm_L\rho^0_L$ and $B^0\to\rho^\mp_L\rho^\pm_L$
branching ratios are not either, since they involve the larger
color-allowed tree amplitude.

Another indication against the resolution in \cite{BY05} is given
by the direct CP asymmetries of the $B\to\pi K$ decays shown in
Table~\ref{tab2}: both $A_{CP}(B^\pm\to \pi^0 K^\pm)$ and
$A_{CP}(B^0\to \pi^\mp K^\pm)$ deviate more from the data, a
consequence expected from the discussion in \cite{LMS05}. As
explained in Sec.~II, the color-suppressed tree amplitude $C'$
needs to be roughly orthogonal to $T'$ in order to have a
vanishing $A_{CP}(B^\pm\to \pi^0 K^\pm)$. The NLO jet function,
though increasing $C'$, does not introduce a large strong phase
relative to $T'$. That is, $C'/T'$ is large, but remains almost
real as in the $B\to\pi\pi$ case \cite{BY05}. This is exactly the
same reason the $B\to\pi K$ puzzle can not be resolved in SCET
\cite{BPS05,WZ0610}: the leading-power SCET formalism demands a
real ratio $C'/T'$, such that a large $C'/T'$ just pushes the SCET
prediction for $A_{CP}(B^\pm\to \pi^0 K^\pm)$, about $-18\%$
\cite{BPS05}, further away from the data. The direct CP asymmetry
of the $B^0\to\pi^0 K^0$ decays, whose tree contribution comes
only from $C'$, is sensitive to the NLO jet function as indicated
in Table~\ref{tab2}. The direct CP asymmetries of the $B\to\pi\pi$
decays, which are tree-dominated, are relatively insensitive to
the NLO jet function.

\section{DISCUSSION}

Before concluding this work, we comment on and compare the various
analyses of the FSI effects in the $B\to\pi K$, $\pi\pi$ decays.
The tiny $B^0\to\pi^0\pi^0$ branching ratio obtained in
perturbative calculations naturally leads to the conjecture that
FSI may play an essential role. Though the estimate of FSI effects
is very model-dependent, the simultaneous applications to
different decay modes can still impose a constraint. The FSI
effects from both the elastic and inelastic channels have been
computed in the Regge model for the $B\to\pi\pi$ decays
\cite{DLLN} and for the $B\to VV$ decays \cite{LLNS}. The
conclusion is that FSI improves the agreement between the
theoretical predictions and the experimental data, but does not
suffice to resolve the $B\to\pi\pi$ puzzle: the $B^0\to\pi^0\pi^0$
branching ratio is increased by FSI only up to 0.1--0.65
\cite{DLLN}. Moreover, the inelastic FSI through the long-distance
charming penguin was found to be negligible in the $B\to\pi\pi$
decays, though it might be important in the $B\to\pi K$ ones. The
reason is that the contribution from the intermediate $D\bar D$
states is CKM suppressed in the former compared to the $D_s\bar D$
states in the latter. This observation differs from that in
\cite{BPRS,BPS05}, where a significant charming-penguin
contribution was claimed. We have pointed out in Sec.~III that the
large charming penguin in \cite{BPRS,BPS05} is a consequence of
fitting the SCET parametrization to the data.

\begin{figure}[tb]
\begin{center}
\includegraphics[scale=0.8]{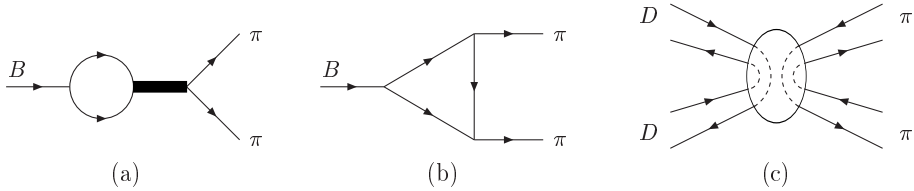}
\caption{Contributions from inelastic FSI. }\label{fig1}
\end{center}
\end{figure}

The inelastic FSI has been also evaluated as the absorptive part
of charmed meson loops shown in Figs.~\ref{fig1}(a) and
\ref{fig1}(b) \cite{CCS}. The two unknown cutoff parameters,
appearing in the form factors associated with the three-meson
vertices, were fixed by the measured $B\to\pi K$ branching ratios.
Note that these parameters should be the same for $B \to \pi K$
and $B \to \pi \pi$ in the SU(3) limit. Applying the same
formalism to the latter, FSI can not resolve the $B\to\pi \pi$
puzzle, even allowing reasonable SU(3) breaking effects for the
cutoff parameters. This result is understandable: the absorptive
amplitudes from Figs.~\ref{fig1}(a) and \ref{fig1}(b) are more or
less orthogonal to the short-distance QCD penguin amplitudes in
the $B\to\pi \pi$ decays, so that their effect is minor. Hence,
the conclusion in \cite{CCS} is the same as in \cite{DLLN}. That
is, the charming penguin is not enough to explain the observed
$B\to\pi\pi$ branching ratios.

Then additional dispersive amplitudes must be taken into account
in \cite{CCS}. Those from Figs.~\ref{fig1}(a) and \ref{fig1}(b),
though calculable in the framework of \cite{CCS}, were not
considered. If considered, they, also contributing to the $B\to\pi
K$ decays, would change the earlier predictions. Therefore, a
brand new mechanism, the dispersive amplitude from the meson
annihilation $D\bar D\to \pi\pi$ shown in Fig.~\ref{fig1}(c), was
introduced. There is no corresponding diagram for the $B\to\pi K$
decays. However, this amplitude is beyond the theoretical
framework, i.e., it can not be expressed in terms of the Feynman
rules derived in \cite{CCS}. The four free parameters, namely, the
two cutoff parameters involved in Figs.~\ref{fig1}(a) and
\ref{fig1}(b), and the real and imaginary contributions from
Fig.~\ref{fig1}(c), were then determined by the four pieces of the
$B\to\pi\pi$ data: the three branching ratios and the direct CP
asymmetry $A_{CP}(B^0\to\pi^\mp\pi^\pm)$. That is, the
$B^0\to\pi^0 \pi^0$ branching ratio has been treated as an input.
The point of \cite{CCS} is to predict the direct CP asymmetries of
the $B^0\to\pi^0 \pi^0$ and $B^\pm\to \pi^\pm\pi^0$ decays, using
the parameters fixed above.

The rescattering among the final states of the $B\to PP$ decays
with $P=\pi$, $K$, and $\eta$ has been studied in \cite{CHY}.
These elastic FSI effects were parameterized in terms of two
strong phases, which, together with the $B\to\pi$ and $B\to K$
form factors, and the chiral enhancement scale, were then
determined by a global fit to the data, including the measured
$B^0\to\pi^0 \pi^0$ branching ratio. Nevertheless, the feature of
the elastic FSI effects, i.e., the correlated decrease and increase
of the $B^0\to\pi^\mp\pi^\pm$ and $B^0\to\pi^0 \pi^0$ branching
ratios, respectively, was noticed \cite{CHY}. A FSI phase
difference between the two $B\to\pi\pi$ isospin amplitudes with
$I=0$, $1$ has been introduced in \cite{KP0601}, which was then
varied to fit the $B\to\pi\pi$ data. Therefore, no explanation for
the large $B^0\to\pi^0 \pi^0$ branching ratio was provided from
the viewpoint of FSI.

There exist other global fits based on different parametrizations
for the charmless $B$ meson decays. For example, the large ratio
$C/T$ was extracted by fitting the quark-amplitude parametrization
to the $B\to\pi\pi$ data
\cite{BFRPR,Y03,Charng,HM04,CGRS,Ligeti04,WZ,BHLD}. No responsible
mechanism was addressed, though the largeness of $C$ was
translated into the largeness of the QCD penguin with an internal
$t$ quark and/or of the exchange amplitudes in \cite{BFRPR}. The
QCDF formalism, in which the twist-3 spectator and annihilation
amplitudes with the end-point singularities were parameterized as
mentioned in Sec.~III, has been implemented into a global fit to
the data \cite{Du,Alek,CWW}. To reach a better fit, the free
parameters involved in QCDF must take different values for the
$B\to PP$, $PV$, $VP$ modes. These parameters have been tuned to
account for the $B\to\pi\pi$ data in \cite{KP0601}. As emphasized
before, the analysis must be also applied to other modes in order
to obtain a consistent picture: the parameters preferred in
\cite{KP0601} lead to a large real $C/T$, which is not favored by
the data of the $B\to\pi K$ direct CP asymmetries as stated in
Sec.~III.

After carefully investigating the proposals available in the
literature, we have found that none of them can really resolve the
$B\to\pi\pi$ puzzle. The NLO PQCD analysis has confirmed that it
is unlikely to accommodate the $B\to\pi\pi$, $\rho\rho$ data
simultaneously (the NLO PQCD predictions are consistent with the
$B\to\rho\rho$ data). The $B\to\pi\pi$ decays have been studied in
the framework of light-cone sum rules (LCSR) \cite{KMMM}, where a
small $B^0\to\pi^0\pi^0$ branching ratio was also observed. Since
there is only little difference between the sum rules for the
$B\to\pi\pi$ and $B\to\rho\rho$ modes, we expect that the
conclusion from LCSR will be the same as from PQCD. The resolution
with the input of the NLO SCET jet function into QCDF \cite{BY05}
does not survive the constraint from the $B\to\rho\rho$ data, and
renders the $B^0\to\pi^\mp K^\pm$ and $B^\pm\to\pi^0 K^\pm$ direct
CP asymmetries deviate more away from the measured values. We
conclude that the $B\to\rho\rho$ data have seriously constrained
the possibility of resolving the $B\to\pi\pi$ puzzle in the
available theoretical approaches.

\vskip 1.0cm

We thank  H.Y. Cheng, C.K. Chua, D. Pirjol, D. Yang, and R. Zwicky
for useful discussions. This work was supported by the National
Science Council of R.O.C. under Grant No. NSC-94-2112-M-001-001,
by the Taipei branch of the National Center for Theoretical
Sciences, and by the U.S. Department of Energy under Grant  No.
DE-FG02-90ER40542.

\appendix

\section{Transverse Helicity Amplitudes}

In this Appendix we present the factorization formulas for the
transverse helicity amplitudes:
\begin{eqnarray}
F_{Ne4}(a)
&=&
16  \pi C_F m_B^2
\int_0^1 dx_1 dx_2 \int_0^{\infty} b_1db_1\, b_2db_2\, \phi_B(x_1,b_1)
\nonumber \\
& &\times
r_3\,
\bigg\{
\left[
  \phi_{2}^T(\overline{x_2}) + 2 r_{2} \phi_{2}^v(\overline{x_2}) + r_{2} x_2
  \left(\phi_{2}^v(\overline{x_2}) + \phi_{2}^a(\overline{x_2}) \right)
\right]
E_{e}(t) h_{e}(A,B,b_1,b_2,x_2)
\nonumber\\
& &\;\;\;\;\;\;\;\;\;\;\;\;
+ r_{2}
  \left(\phi_{2}^v(\overline{x_2}) - \phi_{2}^a(\overline{x_2}) \right)
E_{e}(t') h_{e}(A',B',b_2,b_1,x_1)
\bigg\}\;,
\\
F_{Te4}(a)
&=&
32  \pi C_F m_B^2
\int_0^1 dx_1 dx_2 \int_0^{\infty} b_1db_1\, b_2db_2\, \phi_B(x_1,b_1)
\nonumber \\
& &\times
r_3 \,
\bigg\{
\left[
  \phi_{2}^T(\overline{x_2}) - 2 r_{2} \phi_{2}^a(\overline{x_2}) - r_{2} x_2
  \left(\phi_{2}^v(\overline{x_2}) + \phi_{2}^a(\overline{x_2}) \right)
\right]
E_{e}(t) h_{e}(A,B,b_1,b_2,x_2)
\nonumber\\
& &\;\;\;\;\;\;\;\;\;\;\;\;
+ r_{2}
  \left(\phi_{2}^v(\overline{x_2}) - \phi_{2}^a(\overline{x_2}) \right)
E_{e}(t') h_{e}(A',B',b_2,b_1,x_1)
\bigg\}\;,
\\
%
F_{Ne6}(a) &=& F_{Te6}(a)\ =\ 0 \;,
\\
%
F_{Na4}(a)
&=&
16 \pi C_F m_B^2\, r_{2}r_{3}
\int_0^1 dx_2 dx_3 \int_0^{\infty} b_2db_2\, b_3db_3\,
\nonumber \\
& &\times \bigg\{
\left[
  (1-x_3)
  \left(
    \phi_{2}^v(\overline{x_2}) \phi_{3}^a(\overline{x_3})
    + \phi_{2}^a(\overline{x_2}) \phi_{3}^v(\overline{x_3})
  \right)
  + (1+x_3)
  \left(
    \phi_{2}^v(\overline{x_2}) \phi_{3}^v(\overline{x_3})
    + \phi_{2}^a(\overline{x_2}) \phi_{3}^a(\overline{x_3})
  \right)
\right]
\nonumber \\
& &\;\;\;\;\;\;\times
E_{a}(t) h_{e}(A,B,b_2,b_3,x_3)
\nonumber\\
& &\;\;\;\;\;\;
- \left[
  (2-x_2)
  \left(
    \phi_{2}^v(\overline{x_2}) \phi_{3}^v(\overline{x_3})
    + \phi_{2}^a(\overline{x_2}) \phi_{3}^a(\overline{x_3})
  \right)
  - x_2
  \left(
    \phi_{2}^v(\overline{x_2}) \phi_{3}^a(\overline{x_3})
    + \phi_{2}^a(\overline{x_2}) \phi_{3}^v(\overline{x_3})
  \right)
\right]
\nonumber \\
& &\;\;\;\;\;\;\times
E_{a}(t') h_{e}(A',B',b_3,b_2,x_2)
\bigg\}\;,
\nonumber\\
&=&
16 \pi C_F m_B^2\, r_{2}r_{3}
\int_0^1 dx_2 dx_3 \int_0^{\infty} b_2db_2\, b_3db_3\,
\nonumber \\
& &\times \bigg\{
\left[
  x_3
  \left( \phi_{2}^v(\overline{x_2}) - \phi_{2}^a(\overline{x_2}) \right)
  \left( \phi_{3}^v(\overline{x_3}) - \phi_{3}^a(\overline{x_3}) \right)
  +
  \left( \phi_{2}^v(\overline{x_2}) + \phi_{2}^a(\overline{x_2}) \right)
  \left( \phi_{3}^v(\overline{x_3}) + \phi_{3}^a(\overline{x_3}) \right)
\right]
\nonumber \\
& &\;\;\;\;\;\;\times
E_{a}(t) h_{e}(A,B,b_2,b_3,x_3)
\nonumber\\
& &\;\;\;\;\;\;
- \left[
  (1-x_2)
  \left( \phi_{2}^v(\overline{x_2}) + \phi_{2}^a(\overline{x_2}) \right)
  \left( \phi_{3}^v(\overline{x_3}) + \phi_{3}^a(\overline{x_3}) \right)
  +
  \left( \phi_{2}^v(\overline{x_2}) - \phi_{2}^a(\overline{x_2}) \right)
  \left( \phi_{3}^v(\overline{x_3}) - \phi_{3}^a(\overline{x_3}) \right)
\right]
\nonumber \\
& &\;\;\;\;\;\;\times
E_{a}(t') h_{e}(A',B',b_3,b_2,x_2)
\bigg\}\;,
\\
F_{Ta4}(a)
&=&
-\, 32 \pi C_F m_B^2\, r_{2}r_{3}
\int_0^1 dx_2 dx_3 \int_0^{\infty} b_2db_2\, b_3db_3\,
\nonumber \\
& &\times \bigg\{
\left[
  (1-x_3)
  \left(
    \phi_{2}^v(\overline{x_2}) \phi_{3}^v(\overline{x_3})
    + \phi_{2}^a(\overline{x_2}) \phi_{3}^a(\overline{x_3})
  \right)
  + (1+x_3)
  \left(
    \phi_{2}^v(\overline{x_2}) \phi_{3}^a(\overline{x_3})
    + \phi_{2}^a(\overline{x_2}) \phi_{3}^v(\overline{x_3})
  \right)
\right]
\nonumber \\
& &\;\;\;\;\;\;\times
E_{a}(t) h_{e}(A,B,b_2,b_3,x_3)
\nonumber\\
& &\;\;\;\;\;\;
- \left[
  (2-x_2)
  \left(
    \phi_{2}^v(\overline{x_2}) \phi_{3}^a(\overline{x_3})
    + \phi_{2}^a(\overline{x_2}) \phi_{3}^v(\overline{x_3})
  \right)
  - x_2
  \left(
    \phi_{2}^v(\overline{x_2}) \phi_{3}^v(\overline{x_3})
    + \phi_{2}^a(\overline{x_2}) \phi_{3}^a(\overline{x_3})
  \right)
\right]
\nonumber \\
& &\;\;\;\;\;\;\times
E_{a}(t') h_{e}(A',B',b_3,b_2,x_2)
\bigg\}\;,
\nonumber\\
&=&
32 \pi C_F m_B^2\, r_{2}r_{3}
\int_0^1 dx_2 dx_3 \int_0^{\infty} b_2db_2\, b_3db_3\,
\nonumber \\
& &\times \bigg\{
\left[
  x_3
  \left( \phi_{2}^v(\overline{x_2}) - \phi_{2}^a(\overline{x_2}) \right)
  \left( \phi_{3}^v(\overline{x_3}) - \phi_{3}^a(\overline{x_3}) \right)
  -
  \left( \phi_{2}^v(\overline{x_2}) + \phi_{2}^a(\overline{x_2}) \right)
  \left( \phi_{3}^v(\overline{x_3}) + \phi_{3}^a(\overline{x_3}) \right)
\right]
\nonumber \\
& &\;\;\;\;\;\;\times
E_{a}(t) h_{e}(A,B,b_2,b_3,x_3)
\nonumber\\
& &\;\;\;\;\;\;
+ \left[
  (1-x_2)
  \left( \phi_{2}^v(\overline{x_2}) + \phi_{2}^a(\overline{x_2}) \right)
  \left( \phi_{3}^v(\overline{x_3}) + \phi_{3}^a(\overline{x_3}) \right)
  -
  \left( \phi_{2}^v(\overline{x_2}) - \phi_{2}^a(\overline{x_2}) \right)
  \left( \phi_{3}^v(\overline{x_3}) - \phi_{3}^a(\overline{x_3}) \right)
\right]
\nonumber \\
& &\;\;\;\;\;\;\times
E_{a}(t') h_{e}(A',B',b_3,b_2,x_2)
\bigg\}\;,
\\
%
F_{Na6}(a)
&=&
32 \pi C_F m_B^2
\int_0^1 dx_2 dx_3 \int_0^{\infty} b_2db_2\, b_3db_3\,
\nonumber \\
& &\times
\bigg\{
r_{2} \left(\phi_{2}^v(\overline{x_2}) + \phi_{2}^a(\overline{x_2}) \right)
\phi_{3}^T(\overline{x_3})
E_{a}(t) h_{e}(A,B,b_2,b_3,x_3)
\nonumber\\
& &\;\;\;\;\;\;
+ r_{3} \phi_{2}^T(\overline{x_2})
  \left(\phi_{3}^v(\overline{x_3}) - \phi_{3}^a(\overline{x_3}) \right)
E_{a}(t') h_{e}(A',B',b_3,b_2,x_2)
\bigg\}\;,
\\
F_{Ta6}(a) &=& 2\,F_{Na6}(a)\;,
\\
%
{\cal M}_{Ne4}(a')
&=&
32 \pi C_F \frac{\sqrt{2N_c}}{N_c}m_B^2\, r_3
\int_0^1 dx_1dx_2dx_3 \int_0^{\infty} b_1 db_1\, b_3 db_3\, \phi_B(x_1,b_1)
\nonumber \\
& &\times \bigg\{
(1-x_3) \phi_{2}^T(\overline{x_2})
  \left(\phi_{3}^v(\overline{x_3}) - \phi_{3}^a(\overline{x_3}) \right)
E'_{e}(t) h_n(A,B,b_1,b_3)
\nonumber \\
& &\;\;\;\;\;\;
+ \left[
  x_3 \phi_{2}^T(\overline{x_2})
  \left(\phi_{3}^v(\overline{x_3}) - \phi_{3}^a(\overline{x_3}) \right)
  - 2 r_2 (x_2+x_3)
  \left(
    \phi_{2}^v(\overline{x_2}) \phi_{3}^v(\overline{x_3})
    + \phi_{2}^a(\overline{x_2}) \phi_{3}^a(\overline{x_3})
  \right)
\right]
\nonumber\\
& &\;\;\;\;\;\;\times
E'_{e}(t') h_n(A',B',b_1,b_3)
\bigg\}\;,
\nonumber\\
&=&
32 \pi C_F \frac{\sqrt{2N_c}}{N_c}m_B^2\, r_3
\int_0^1 dx_1dx_2dx_3 \int_0^{\infty} b_1 db_1\, b_3 db_3\, \phi_B(x_1,b_1)
\nonumber \\
& &\times \bigg\{
(1-x_3) \phi_{2}^T(\overline{x_2})
  \left(\phi_{3}^v(\overline{x_3}) - \phi_{3}^a(\overline{x_3}) \right)
E'_{e}(t) h_n(A,B,b_1,b_3)
\nonumber \\
& &\;\;\;\;\;\;
+ \left[
  x_3 \phi_{2}^T(\overline{x_2})
  \left(\phi_{3}^v(\overline{x_3}) - \phi_{3}^a(\overline{x_3}) \right)
  - r_2 (x_2+x_3)
    \left( \phi_{2}^v(\overline{x_2}) + \phi_{2}^a(\overline{x_2}) \right)
    \left( \phi_{3}^v(\overline{x_3}) + \phi_{3}^a(\overline{x_3}) \right)
  \right.\nonumber\\
  & &\left.\;\;\;\;\;\;\;\;\;\;\;
  - r_2 (x_2+x_3)
    \left( \phi_{2}^v(\overline{x_2}) - \phi_{2}^a(\overline{x_2}) \right)
    \left( \phi_{3}^v(\overline{x_3}) - \phi_{3}^a(\overline{x_3}) \right)
\right]
E'_{e}(t') h_n(A',B',b_1,b_3)
\bigg\}\;,
\\
{\cal M}_{Te4}(a')
&=&
64 \pi C_F \frac{\sqrt{2N_c}}{N_c}m_B^2\, r_3
\int_0^1 dx_1dx_2dx_3 \int_0^{\infty} b_1 db_1\, b_3 db_3\, \phi_B(x_1,b_1)
\nonumber \\
& &\times
\bigg\{
  (1-x_3) \phi_{2}^T(\overline{x_2})
  \left(\phi_{3}^v(\overline{x_3}) - \phi_{3}^a(\overline{x_3}) \right)
E'_{e}(t) h_n(A,B,b_1,b_3)
\nonumber \\
& &\;\;\;\;\;\;
+ \left[ x_3 \phi_{2}^T(\overline{x_2})
  \left(\phi_{3}^v(\overline{x_3}) - \phi_{3}^a(\overline{x_3}) \right)
  + 2 r_2 (x_2+x_3)
  \left(
    \phi_{2}^v(\overline{x_2}) \phi_{3}^a(\overline{x_3})
    + \phi_{2}^a(\overline{x_2}) \phi_{3}^v(\overline{x_3})
  \right)
\right]
\nonumber \\
& &\;\;\;\;\;\;\times
E'_{e}(t') h_n(A',B',b_1,b_3)
\bigg\}\;,
\nonumber\\
&=&
64 \pi C_F \frac{\sqrt{2N_c}}{N_c}m_B^2\, r_3
\int_0^1 dx_1dx_2dx_3 \int_0^{\infty} b_1 db_1\, b_3 db_3\, \phi_B(x_1,b_1)
\nonumber \\
& &\times
\bigg\{
  (1-x_3) \phi_{2}^T(\overline{x_2})
  \left(\phi_{3}^v(\overline{x_3}) - \phi_{3}^a(\overline{x_3}) \right)
E'_{e}(t) h_n(A,B,b_1,b_3)
\nonumber \\
& &\;\;\;\;\;\;
+ \left[ x_3 \phi_{2}^T(\overline{x_2})
  \left(\phi_{3}^v(\overline{x_3}) - \phi_{3}^a(\overline{x_3}) \right)
  + r_2 (x_2+x_3)
    \left( \phi_{2}^v(\overline{x_2}) + \phi_{2}^a(\overline{x_2}) \right)
    \left( \phi_{3}^v(\overline{x_3}) + \phi_{3}^a(\overline{x_3}) \right)
  \right.\nonumber\\
  & &\left.\;\;\;\;\;\;\;\;\;\;\;
  - r_2 (x_2+x_3)
    \left( \phi_{2}^v(\overline{x_2}) - \phi_{2}^a(\overline{x_2}) \right)
    \left( \phi_{3}^v(\overline{x_3}) - \phi_{3}^a(\overline{x_3}) \right)
\right]
E'_{e}(t') h_n(A',B',b_1,b_3)
\bigg\}\;,
\\
%
{\cal M}_{Ne5}(a')
&=&
- {\cal M}_{Ne4}(a')
\;,
\\
{\cal M}_{Te5}(a')
&=&
- {\cal M}_{Te4}(a')
\;,
\\
%
{\cal M}_{Ne6}(a')
&=&
-\, 32  \pi C_F \frac{\sqrt{2N_c}}{N_c}m_B^2\, r_{2}
\int_0^1 dx_1dx_2dx_3\int_0^{\infty} b_1 db_1\, b_3 db_3\,\phi_B(x_1,b_1)
\nonumber \\
& &\times
x_2 \left(\phi_{2}^v(\overline{x_2}) + \phi_{2}^a(\overline{x_2}) \right)
\phi_{3}^T(\overline{x_3})
\left[
  E'_{e}(t) h_n(A,B,b_1,b_3)
  + E'_{e}(t') h_n(A',B',b_1,b_3)
\right]\;,
\\
{\cal M}_{Te6}(a') &=& 2\, {\cal M}_{Ne6}(a')\;,
\\
%
{\cal M}_{Na4}(a')
&=&
-\, 64 \pi C_F \frac{\sqrt{2N_c}}{N_c}m_B^2\, r_{2} r_{3}
\int_0^1 dx_1dx_2dx_3 \int_0^{\infty} b_1 db_1\, b_3 db_3\,\phi_B(x_1,b_1)
\nonumber \\
& &\times
\left(
  \phi_{2}^v(\overline{x_2}) \phi_{3}^v(\overline{x_3})
  + \phi_{2}^a(\overline{x_2}) \phi_{3}^a(\overline{x_3})
\right)
E'_{a}(t') h_n(A',B',b_3,b_1)\;,
\nonumber\\
&=&
-\, 32 \pi C_F \frac{\sqrt{2N_c}}{N_c}m_B^2\, r_{2} r_{3}
\int_0^1 dx_1dx_2dx_3 \int_0^{\infty} b_1 db_1\, b_3 db_3\,\phi_B(x_1,b_1)
\nonumber \\
& &\times
\bigg\{
  \left( \phi_{2}^v(\overline{x_2}) + \phi_{2}^a(\overline{x_2}) \right)
  \left( \phi_{3}^v(\overline{x_3}) + \phi_{3}^a(\overline{x_3}) \right)
  +
  \left( \phi_{2}^v(\overline{x_2}) - \phi_{2}^a(\overline{x_2}) \right)
  \left( \phi_{3}^v(\overline{x_3}) - \phi_{3}^a(\overline{x_3}) \right)
\bigg\}
\nonumber \\
& &\;\;\;\;\;\;\times
E'_{a}(t') h_n(A',B',b_3,b_1)\;,
\\
{\cal M}_{Ta4}(a')
&=&
128 \pi C_F \frac{\sqrt{2N_c}}{N_c}m_B^2\, r_{2} r_{3}
\int_0^1 dx_1dx_2dx_3 \int_0^{\infty} b_1 db_1\, b_3 db_3\,\phi_B(x_1,b_1)
\nonumber \\
& &\times
\left(
  \phi_{2}^v(\overline{x_2}) \phi_{3}^a(\overline{x_3})
  + \phi_{2}^a(\overline{x_2}) \phi_{3}^v(\overline{x_3})
\right)
E'_{a}(t') h_n(A',B',b_3,b_1)\;,
\nonumber\\
&=&
64 \pi C_F \frac{\sqrt{2N_c}}{N_c}m_B^2\, r_{2} r_{3}
\int_0^1 dx_1dx_2dx_3 \int_0^{\infty} b_1 db_1\, b_3 db_3\,\phi_B(x_1,b_1)
\nonumber \\
& &\times
\bigg\{
  \left( \phi_{2}^v(\overline{x_2}) + \phi_{2}^a(\overline{x_2}) \right)
  \left( \phi_{3}^v(\overline{x_3}) + \phi_{3}^a(\overline{x_3}) \right)
  -
  \left( \phi_{2}^v(\overline{x_2}) - \phi_{2}^a(\overline{x_2}) \right)
  \left( \phi_{3}^v(\overline{x_3}) - \phi_{3}^a(\overline{x_3}) \right)
\bigg\}
\nonumber \\
& &\;\;\;\;\;\;\times
E'_{a}(t') h_n(A',B',b_3,b_1)\;,
\\
%
{\cal M}_{Na5}(a') &=& {\cal M}_{Na4}(a')\;,
\\
{\cal M}_{Ta5}(a') &=& {\cal M}_{Ta4}(a')\;,
\\
%
{\cal M}_{Na6}(a')
&=&
32 \pi C_F \frac{\sqrt{2N_c}}{N_c}m_B^2
\int_0^1 dx_1dx_2dx_3 \int_0^{\infty} b_1 db_1\, b_3 db_3\,\phi_B(x_1,b_1)
\nonumber \\
& &\times
\bigg\{
\left[
  r_3 x_3 \phi_{2}^T(\overline{x_2})
  \left(\phi_{3}^v(\overline{x_3}) - \phi_{3}^a(\overline{x_3}) \right)
  - r_2 (1-x_2)
  \left(\phi_{2}^v(\overline{x_2}) + \phi_{2}^a(\overline{x_2}) \right)
  \phi_{3}^T(\overline{x_3})
\right]
\nonumber \\
& &\;\;\;\;\;\;\times
E'_{a}(t) h_n(A,B,b_3,b_1)
\nonumber \\
& &\;\;\;\;\;\;
+ \left[
  r_3 (2-x_3) \phi_{2}^T(\overline{x_2})
  \left(\phi_{3}^v(\overline{x_3}) - \phi_{3}^a(\overline{x_3}) \right)
  - r_2 (1+x_2)
  \left(\phi_{2}^v(\overline{x_2}) + \phi_{2}^a(\overline{x_2}) \right)
  \phi_{3}^T(\overline{x_3})
\right]
\nonumber \\
& &\;\;\;\;\;\;\times
E'_{a}(t') h_n(A',B',b_3,b_1)
\bigg\}\;,
\\
{\cal M}_{Ta6}(a') &=& 2\, {\cal M}_{Na6}(a')\;.
%
\end{eqnarray}
The quark-loop corrections ${\cal M}^{(q)}_{N,T}$ for $q=u$, $c$,
and $t$, and the magnetic-penguin corrections ${\cal
M}^{(g)}_{N,T}$ to the transverse components are written as
\begin{eqnarray}
{\cal M}^{(q)}_{N} &=& - 16  m_B^2 \frac{C_F^2}{\sqrt{2N_c}}\, r_3
\int_0^1 dx_1 dx_2 dx_3\int_0^{\infty} b_1db_1 b_2db_2
\phi_B(x_1,b_1)
\nonumber \\
& & \times \big\{\left[
  \phi_2^T(\overline{x_2})
  \left(\phi_{3}^v(\overline{x_3}) + \phi_{3}^a(\overline{x_3}) \right)
    \right. \nonumber \\
    & &
\;\;\;\;\;\;\;\;\;\;\; \left.
  + r_2 (2+x_2)
    \phi_{2}^v(\overline{x_2})\phi_{3}^v(\overline{x_3})
  + r_2 x_2
    \phi_{2}^a(\overline{x_2})\phi_{3}^v(\overline{x_3})
  + 4 r_2
    \phi_{2}^a(\overline{x_2})\phi_{3}^a(\overline{x_3})
\right]\nonumber \\
& & \;\;\;\;\;\; \times E^{(q)}(t_q,l^2) h_{e}(A,B,b_1,b_2,x_2)
\nonumber\\
& & \;\;\;\;\;\; +\,
    r_2 \phi_{2}^v(\overline{x_2}) \phi_{3}^v(\overline{x_3})
E^{(q)}(t_q^{\prime},l^{\prime 2}) h_{e}(A',B',b_2,b_1,x_1) \big\}
\;,
\\
{\cal M}^{(q)}_{T} &=& 0 \;,
\\
{\cal M}^{(g)}_{N} &=& 16 m_B^4  \frac{C_F^2}{\sqrt{2N_c}}
\int_0^1 dx_1dx_2dx_3 \int_0^{\infty} b_1db_1\, b_2db_2\,
b_3db_3\, \phi_B(x_1,b_1)
\nonumber \\
& &\times\, \Big\{ \left[ - r_2 (1-x_2^2)
  \left(\phi_{3}^v(\overline{x_3}) + \phi_{3}^a(\overline{x_3}) \right)
  \phi_{3}^T(\overline{x_3})
- r_3 (1+x_2) x_3
  \phi_{2}^T(\overline{x_2})
  \left(\phi_{3}^v(\overline{x_3}) - \phi_{3}^a(\overline{x_3}) \right)
  \right.\nonumber\\
  & &\left.\;\;\;\;\;\;\;\;\;\;\;
- r_2 r_3 (1-x_2)
  \left( \phi_{2}^v(\overline{x_2}) + \phi_{2}^a(\overline{x_2}) \right)
  \left( \phi_{3}^v(\overline{x_3}) + \phi_{3}^a(\overline{x_3}) \right)
  \right.\nonumber\\
  & &\left.\;\;\;\;\;\;\;\;\;\;\;
- r_2 r_3 x_3(1-2x_2)
  \left( \phi_{2}^v(\overline{x_2}) - \phi_{2}^a(\overline{x_2}) \right)
  \left( \phi_{3}^v(\overline{x_3}) - \phi_{3}^a(\overline{x_3}) \right)
\right]
E_{g}(t_q) h_g(A,B,C,b_1,b_2,b_3,x_2)
\nonumber\\
& & \;\;\;\;\;\; - r_2 r_3 x_3
  \left( \phi_{2}^v(\overline{x_2}) - \phi_{2}^a(\overline{x_2}) \right)
  \left( \phi_{3}^v(\overline{x_3}) - \phi_{3}^a(\overline{x_3}) \right)
E_{g}(t_q^{\prime}) h_g(A',B',C',b_2,b_1,b_3,x_1) \Big\} \;,
\\
{\cal M}^{(g)}_{T} &=& 32 m_B^4  \frac{C_F^2}{\sqrt{2N_c}}
\int_0^1 dx_1dx_2dx_3 \int_0^{\infty} b_1db_1\, b_2db_2\,
b_3db_3\, \phi_B(x_1,b_1)
\nonumber \\
& &\times\, \Big\{ \left[ - r_2 (1-x_2^2)
  \left(\phi_{3}^v(\overline{x_3}) + \phi_{3}^a(\overline{x_3}) \right)
  \phi_{3}^T(\overline{x_3})
- r_3 (1+x_2) x_3
  \phi_{2}^T(\overline{x_2})
  \left(\phi_{3}^v(\overline{x_3}) - \phi_{3}^a(\overline{x_3}) \right)
  \right.\nonumber\\
  & &\left.\;\;\;\;\;\;\;\;\;\;\;
+ r_2 r_3 (1-x_2)
  \left( \phi_{2}^v(\overline{x_2}) + \phi_{2}^a(\overline{x_2}) \right)
  \left( \phi_{3}^v(\overline{x_3}) + \phi_{3}^a(\overline{x_3}) \right)
  \right.\nonumber\\
  & &\left.\;\;\;\;\;\;\;\;\;\;\;
- r_2 r_3 x_3(1-2x_2)
  \left( \phi_{2}^v(\overline{x_2}) - \phi_{2}^a(\overline{x_2}) \right)
  \left( \phi_{3}^v(\overline{x_3}) - \phi_{3}^a(\overline{x_3}) \right)
\right]
E_{g}(t_q) h_g(A,B,C,b_1,b_2,b_3,x_2)
\nonumber\\
& &
\;\;\;\;\;\; - r_2 \left[ (1-x_2)
    \phi_{2}^v(\overline{x_2})\phi_{3}^T(\overline{x_3})
- r_3 (1-2x_2)
    \phi_{2}^v(\overline{x_2})\phi_{3}^v(\overline{x_3})
- r_3
    \phi_{2}^v(\overline{x_2})\phi_{3}^a(\overline{x_3})
  \right.\nonumber\\
  & &\left.\;\;\;\;\;\;\;\;\;\;\;
- r_3
    \phi_{2}^a(\overline{x_2})\phi_{3}^v(\overline{x_3})
+ r_3
    \phi_{2}^a(\overline{x_2})\phi_{3}^a(\overline{x_3})
\right]
E_{g}(t_q^{\prime}) h_g(A',B',C',b_2,b_1,b_3,x_1) \Big\} \;.
\end{eqnarray}
The definitions of all the variables and the convolution factors
in the above expressions are referred to \cite{LMS05}.

\end{document}